\documentclass{PoS}

\title{Nuclei of Seyfert galaxies and QSOs - Central engine \& conditions of star formation\\
{\Large Workshop summary and open questions } }

\ShortTitle{Seyfert 2012}

\author{M\'onica Valencia-S.\\ 
		I. Physikalisches Institut, Universit\"at zu K\"oln, Z\"ulpicher Str. 77, 50937 K\"oln, Germany\\
		E-mail: \email{zuther@ph1.uni-koeln.de}}
      
		\author{Jens Zuther\\
        I. Physikalisches Institut, Universit\"at zu K\"oln, Z\"ulpicher Str. 77, 50937 K\"oln, Germany\\
        E-mail: \email{zuther@ph1.uni-koeln.de}}

        \author{Mariangela Vitale\\
        I. Physikalisches Institut, Universit\"at zu K\"oln, Z\"ulpicher Str. 77, 50937 K\"oln, Germany\\
        E-mail: \email{vitale@ph1.uni-koeln.de}}

		\author{Andreas Eckart\\
        I. Physikalisches Institut, Universit\"at zu K\"oln, Z\"ulpicher Str. 77, 50937 K\"oln, Germany\\
        Max-Planck Institut f\"ur Radioastronomie, Auf dem H\"ugel 69, 53121 Bonn, Germany\\
        E-mail: \email{eckart@ph1.uni-koeln.de}}

		\author{Mohammad Zamaninasab\\
        Max-Planck Institut f\"ur Radioastronomie, Auf dem H\"ugel 69, 53121 Bonn, Germany\\
        E-mail: \email{zamanina@mpifr-bonn.mpg.de}}

\abstract{Observationally established correlations between black hole mass and host galaxy structural/dynamical properties (like the $M_{\rm BH}-\sigma_{\ast}$ relation) give support to the idea of an intimate link between the growth of black holes (BHs) and their host galaxies. Active galactic nuclei represent a poorly understood phase (or phases) in the life of a galaxy, during which the BH growth is directly observable.
With the advent of wide-field surveys and high angular resolution instruments, it is now possible to
observe the source of energy in Seyfert galaxies and QSOs, and conduct both statistical and detailed on-object studies.
Combining these two perspectives gives rise to new research questions and methods. This Workshop, 
aimed at discussing these questions and gathering the astronomical community working on AGN, their feeding and feedback mechanisms, and their relations to the host galaxies. 
The purpose of this summary paper is to condense in a few pages some of the ideas and discussion points that were considered during the Workshop. We would also like to call the attention to some open questions that are still matter of debate and drive the research efforts in the field.}

\FullConference{Nuclei of Seyfert galaxies and QSOs - Central engine \& conditions of star formation,\\
                November 6-8, 2012\\
                Max-Planck-Insitut f\"ur Radioastronomie (MPIfR), Bonn, Germany}

\usepackage{url}
\begin{document}

\section{Indroduction}

The workshop ``Nuclei of Seyfert galaxies and QSOs: Central engine and conditions of star formation'' was
held at the Max-Planck-Insitut f\"ur Radioastronomie (MPIfR) in Bonn (Germany). Ninety participants, working at 47 different institutes in 23 countries around the world, met to discuss about the state-of-the-art of the research on AGN and the interplay between the host galaxies and their central  black hole (BH).
The advent of high angular resolution instruments working in the entire electromagnetic spectral range (adaptive-optics assisted integral field spectroscopy like ESO-SINFONI and Gemini NIFS, and very long baseline interferometry such as VLBI), as well as massive panchromatic galaxy surveys (e.g., AKARI all-sky survey, COSMOS, SDSS) opens new perspectives and fosters our understanding of the galaxy/AGN population and cosmic evolution.
Topics ranged from the closest supermassive BHs (including the Galactic Center) out to star formation in high-redshift AGN.
The Workshop sessions were organized as follows:
\begin{enumerate}
\item Nearby galaxies and the Galactic Center
\item Narrow-line Seyfert 1 galaxies
\item Outflows, jets, and feedback
\item Seyfert galaxies and QSOs: Black-hole masses, scaling relations, and star formation
\item Galaxy mergers on all scales
\item Star formation and accretion at high redshift
\end{enumerate}

This paper contains a short summary of the Workshop sessions. This summary does not intend to cover all topics discussed at the meeting, but only to point out some of the interesting discussions, including discrepant ideas, open questions, and new observational data presented at the Workshop.
We draw special attention to the controversial and still not-well-understood phenomena, whose study is  possible only now thanks to the current high angular resolution instrumentation, ${\rm methodologically}$ more complete data surveys, and the development of new methods to analyze archival data.

\section{Disscusion}
{\bf Nearby Galaxies and the Galactic Center}\\
The first session focussed on studies of
nearby galaxies and the quest for ever higher angular resolution - achievable in the radio regime, or in the infrared by interferometric methods - tracing regions on the scale of a few pc, and allowing us to resolve the molecular torus in the nearest AGN. Nearby galaxies are the best targets for learning about the environment of BHs. Observations of the nuclei of host galaxies can hint at possible feeding mechanisms that trigger accretion at different levels of BH activity.

Some authors have opted for a multiwavelegth and statistical approach, while others study particular sources in detail.  M.~Vitale showed, for example, how optical diagnostic diagrams can be successfully used to classify sources with radio and optical emission detected in the FIRST and SDSS surveys. She showed that most of the radio LINERs might be powered by low-luminosity AGN (LLAGN). Another powerful technique for identifying and classifying AGN using multiwavelength photometry was presented by X.~Dong. In the review talk on `Galaxies and their nuclei', C. Mundell stressed the importance of exploring the inner kiloparsec of active galaxies. She presented evidences for short and probably repetitive duty cycles in some AGN, and introduced a new -- statistically significant -- sample of galaxies with integral field spectoscopic (IFU) optical observations, which will allow detailed studies of gas and stellar kinematics in the nucleus of these sources. A. Prieto pointed out that in order to get a proper spectral energy distribution (SED) of an AGN it is necessary to dissect the inner cores of the hosts to resolutions of few tens of parsecs. She also showed some examples of SEDs obtained with observations assisted by adaptive optics in the near-infrared (NIR) and using interferometric techniques in mid-infrared and radio.

Interesting specific cases of nearby AGN observed in the optical or NIR with IFU techniques were also shown. M.~Bremer found that NGC~5850 possesses a counter-rotating disk of gas and that its LINER-like emission is likely not associated to a LLAGN but to shocks in the inter-stellar medium (ISM). In NGC~7172, optically classified as Seyfert 2 with no-hidden polarized broad lines, S. Smajic reported the detection of broad Pa$\alpha$ and Br$\gamma$. In NGC~4151, using the mapped kinematics of the gas in the narrow line region (NLR), C. Iserlohe showed that acceleration takes place in the inner arcsecond ($\sim 70$\,pc) around the nucleus and that the jet is not an evident source of excitation of the narrow line emission.

Current results on the closest galactic nucleus, the Galactic Center (GC), were presented, in particular the potential gas/dust cloud currently (2013-2014) approaching the GC. The nature of this source is still controvertial. M. Schartmann described two possible scenarios for the evolution of the G2/DSO object based on hydrodynamical simulations. He also showed how NIR IFU observations of the evolution of the source can help in discriminating between these possible explanations. A. Eckart, in turn, showed the first K-band identifications of G2/DSO, indicating that it might be a dust-enshrouded star in contrast to a pure gas and dust object. He also presented SgrA*, the Milky Way SMBH, as a paradigm of a LLAGN. Its mass, derived from the motions of the detected S-cluster stars, is most accurately known. N. Sabha found a new contribution to the precession of the orbit of S2  that will enable us to constrain the mass and number of sources located even closer to the BH. R. Capuzzo proposed the scenario of merging of orbitally decaying globular clusters for the formation of this central cluster. The activity of SgrA* is evident in the short-term NIR variability of the total intensity and polarization parameters, as shown by B. Shahzamanian. The GC environment also offers unexpected phenomena; for example, J. Moultaka discovered the presence of CO-ices in the central kiloparsec, implying very low temperatures near to the BH.

\bigskip

{\bf Narrow Line Seyfert 1 galaxies}\\
The second session was dedicated to Narrow-Line Seyfert 1 (NLSy1) galaxies. The properties of NLSy1s place them at one extreme of the distribution of Type 1 AGN in the eigenvector~1 (EV1) diagrams. This has been interpreted in terms of the high Eddington ratio of these sources. Yet, D. Xu showed that the density of the gas in the narrow line region also constitutes a key element in the EV1. This strong connection between BH and galaxy host in NLSy1s opens new routes for understanding the galaxy-AGN co-evolution. On this respect, G. Orban de Xivry discussed evidence for predominant secular evolution of the NLSy1 hosts, in contrast to Broad Line ${\rm Seyfert~1}$ galaxies. One of the most intriguing  properties of NLSy1s is the relatively strong iron emission. A. Rodriguez-Ardila presented a NIR Fe{\sc ii} template constructed from the observations of IZw1 (which complements those in the optical and UV), and confirmed the importance of fluorescence in the excitation of these lines. D. Ilic drove the attention to the variability of the Fe{\sc ii} and hydrogen-emission lines from long-term optical monitoring of Type 1 sources (Ark 564, 3C390.3, and NGC~4151) calling for caution when deriving black hole masses from single-epoch spectroscopic observations.

Especially in the radio and gamma-ray regimes, recent discoveries of relativistic jets in a fraction of NLSy1s are remarkable. The question on the nature of the physical drivers is still open. In the review talk on `Powerful relativistic Jets in Narrow Line Seyfert 1 galaxies', L. Foschini presented a compilation of about 50 radio-loud NLSy1s. Given the low black hole masses of these sources ($M_{\rm BH}\sim10^6 - 10^8 M_{\odot}$), they can be considered the `missing piece of unification of jets at all scales'. Comparing the disk with the jet power of AGN and compact accreting stellar objects, radio-loud NLSy1s would be analogous to neutron stars at Galactic scales. Whether the jet launching is related to the spin of the compact object is not known.
M. Ward and C. Done discussed a new technique to estimate the spin of the black hole in AGN that is not based on the somehow controversial profile of the iron K-alpha line, but on the shape of the overall high energy (optical to X-rays) spectrum.
M. Valencia-S. addressed the question whether NLSy1s can be considered a distinct class of AGN. Using the galaxy IRAS~01072+4954 as a case study, and other sources from the literature, she argued that NLSy1s are only the low black hole mass tail of the Type 1 population. V. Cracco presented a low-redshift sample of NLSy1s drawn from the SDSS that can be used to study the nature of these sources in the frame of the unified model.

\bigskip

{\bf Outflows, Jets and Feedback}\\
Session 3 was devoted to different forms of feedback observed in low- and high-redshift AGN, such as jets and outflows. Feedback appears to play a major role in the structure formation and galaxy evolution. Two flavors of feedback are suggested in the literature: the `radio mode', associated with radio jets/lobes heating the gas, and the `quasar mode', where high-to-medium velocity outflows are powered by highly accreting objects. However, in her review talk `Outflows, feedback and jets', R. Morganti showed evidences of both mechanisms been in place in several objects and argued that their efficiency and impact on the host galaxy depend on the coupling with the interstellar medium (ISM) from parsec to kpc scales. She presented examples of sources (e.g. IC~5067 and 3C293) where massive molecular outflows are most likely generated by the interaction between the jet and the ISM.
Outflows can also be used to look for signs of former AGN activity as presented by A. Shulevski. He showed that multi-frequency radio observations, obtained e.g. with LOFAR and Westerbork, deliver important constraints on the age of the extended outflow/jet structures that can be related to active and inactive periods of the AGN.

Significant amounts of  highly ionized gas seem to be launched closer to the AGN, however the physical mechanisms and possible connection between them are still matter of debate. In the case of ultra-fast outflows observed in X-rays, J.M. Ramirez found a relation between the bolometric luminosity of the AGN and the velocity of the outflow, which is produced via line-radiation pressure.
Conversely, F. Hamann argued that using this scenario to describe UV-outflows would require some kind of shielding to prevent over-ionization of the gas. Based on the absence of indications for such shielding in a sample of quasars with mini-broad absorption lines (BALs), he proposed the wind ionization to be controlled by the high gas density instead. E. Cooper presented the impressive case of WPVS007, a NLSy1 galaxy with BALs. In the past 15 years, the velocity of the UV-outflow in this source has increased by a factor of ten, as it can be seen from the deepening and broadening of the absorption features in the HST spectra taken in 1996, 2007, and 2010.
Ionized outflows with velocities of few hundreds of ${\rm km/s}$ are not only observed in AGN but also in Starburst galaxies. A. Tsai presented the cases of NGC~2146 and NGC~3628, two edge-on starburst galaxies with molecular outflows observed using the Nobeyama Millimeter Array.

The conditions for jet launching (e.g., in the accretion flow, spin of the BH, intensity of the magnetic field) are still unknown. J.A. Fernandez Ontiveros showed that, in a sample of nearby LLAGN (LINER), high angular resolution SEDs confirm the absence of the blue-bump (predicted for radiatively inefficient accretion disks), show no signs of a mid-IR bump (expected when the torus is present), and suggest all spectral bands to be purely dominated by jet emission. In contrast, the radio (from EVN) and X-ray emission of the Seyfert 2s studied by Z. Paragi is consistent with an ADAF-jet scenario, where the jet dominates the emission at high energies. H. Flohic found that in comparison to the QSO SEDs, the mean LLAGN LINER SEDs have X-ray band slopes similar to radio-quiet AGN, while the radio luminosities and slopes are similar to those of radio-loud quasars.
In blazars (BL Lacs and Flat Spectrum Radio Quasars, FSRQs) the jet emission dominates at all wavelengths, because of the fortuitous alignment of the jet axis and the line of sight towards the BH. Many blazars also emit $\gamma$-rays, but the production mechanism is still unclear. L. Fuhrmann reported a positive correlation between the flux variations at radio and $\gamma$-rays in stacked spectra of 56 blazars. Radio multi-frequency monitoring of the $\gamma$-ray emitting blazar S5~0716+714 was presented by B. Rani. She found that the variations of the spectral index that occur during flaring events might be explained by variations in the geometry of the jet (twists or helical structures). Furthermore, E. Angelakis presented a similar study of 4 NLSy1s with $\gamma$-ray emission detected also by FERMI. He found that the radio light curves and linear polarization of these sources are similar to those typically observed in blazars.

\bigskip

{\bf Seyferts and QSOs: BH masses, scaling relations and star formation}\\
Major themes of Session 4 were the estimation of black-hole masses, the AGN toroidal obscuration, the scaling relations between BHs and host galaxies, star formation and AGN relations, and new surveys and observational techniques.

B. Peterson adressed the implicit assumptions and reliability of the reverberation mapping (RM) technique for determining black hole masses. He also showed that in the handful of sources where the velocity resolved RM has been applied, the broad line region (BLR) geometry displays a disk-like structure and signs of infall from the hydrogen-emission lines. In this respect, J. Runnoe showed evidences for orientation-dependent H$\beta$ profiles, which can change the estimated black hole masses by $\sim1$ dex in the worst-case scenario. According to her analysis, this can be prevented using the C{\sc iv} line as a probe for the BLR kinematics. M. Zetzl also showed that the profiles of H$\beta$, HeII, and C{\sc iv} broad lines can be modelled by Lorentzian profiles, caused by turbulence, broadened by rotation.
F. Pozo described an innovate photometric technique for RM, which uses only a couple of broad- and one narrow-band filters. Using the flux-variation gradient-method, he demonstrated that it is possible to efficiently remove the galaxy contribution and obtain accurate BLR sizes and AGN luminosities. Moreover, applying this technique to study the BLR geometry in 3C~120, C. Bruckmann found that a disk-like configuration of the BLR can explain the sharp features observed in the H$\beta$ light curve.

A key element in the unified model is the torus, whose nature and appearance are still unknown. Combining data from the WISE catalogue and the SDSS spectroscopic survey, Y. Toba showed indications for the covering factor of the dusty torus to decrease with the mid-infrared (MIR) luminosity. D. Asmus presented a MIR AGN atlas containing approx. 250 sources with single-dish high-resolution photometrical information. He also confirmed the ${\rm MIR - X}$-ray correlation in Seyferts and LINERs, which extends over six orders of magnitude. On the other hand, from high-angular resolution infrared interferometry of the tori of nearby AGN, K. Tristram reported the presence of a two-component structure: one in a disk-like geometry, and the other within the ionization cone. Further surprising results were presented by J. Polednikova, who reported the discovery of short-term (timescale of hours) optical variability in Type 2 quasars. 

A large compilation of black hole masses of quiescent galaxies (72) obtained via dynamical methods and AGN (25) using reverberation mapping was presented by J.-H. Woo. He showed that the $M_{\rm BH} - \sigma_{\ast}$ relations of active and inactive galaxies are consistent with each other, and that the small difference in fit-slopes can be attributed to the inherent selection effects of both samples. For example, RM estimations are only applicable when the AGN is sufficiently bright (i.e. bias against low $M_{\rm BH}$), with a measurable variability amplitude (i.e. strong bias against massive BHs).
These selection biases can harm the conclusions about the relation between the BH and the host galaxy and its evolution if they are not well accounted for. In the review talk on `Demographics of QSOs and their hosts', L.~Wisotzki extensively discussed the importance of improving the knowledge on the distribution functions of the host galaxies and AGN properties, in order to constraint these biases better. 

It is matter of debate whether the same kind of galaxies host active and quiet BHs, or Type 1 and Type 2 AGN; whether they constitute phases of some evolutionary sequence or not; and what the role of the star formation and AGN feedback is in those scenarios. Using infrared AKARI data, T. Miyaji and A. Castro found no significant differences between star-formation luminosity in Type 1 and Type 2 AGN. However, B. Villarroel found that the morphology and environment of Type 1 and Type 2 AGN are different. 
Comparing the emission-line and stellar-population properties of thousands of objects from the SDSS, G. La Mura found that there are various sequences connecting star forming galaxies, Seyfert 2s, and normal galaxies and that they might indicate some evolutionary relation.
Another kind of evolutionary scenario was proposed by P.-C. Yu, who found indications for older stellar populations in non-Hidden Broad Line Region (non-HBLR) Seyfert 2 galaxies than in HBLR Seyfert 2s. He suggested that Sy1s and HBLR Sy2s might respectively evolve into unabsorbed and absorbed non-HBLR Sy2 objects.
M. Karouzos studied a sample of AKARI radio-AGN, and found that radio-loud sources at $z\lesssim2$ have specific star formation rates  consistent with those expected for galaxies at those redshifts, implying that the radio-mode feedback does suppress, but not fully quench, the star formation in these objects.  Similarly, D. Dicken reported the detection of recent star-formation in only $\sim 35$\% of radio-selected ($S_{\rm 2.7\,GHz}>2$\,Jy) local AGN ($z<0.7$). 
L. Popovi\'c presented a new Fe{\sc ii} template to model the optical spectra in Type 1 AGN. He found that the iron emission in star-formation dominated Type 1s (i.e. with $\log([$O{\sc iii}$]/{\rm H}\beta_{\rm narrow})<0.5$) is somewhat stronger than in AGN-dominated objects.
S. Raimundo showed high-angular resolution NIR studies of the central 150 pc of the NLSy1 galaxy MCG-06-30-15, where she found a counter-rotating stellar core that could be the result of a 65-Myr old starburst episode. 

B. Husemann announced the first public data release of the Calar Alto Legacy Integral Field Area (CALIFA) survey, which contains low- and medium- resolution IFU optical data of hundreds of nearby galaxies ($0.005<z<0.03$). With the aim of gaining angular resolution in the optical, L. Labadie explored the capabilities of the next-to-come Adaptive Optics Lucky Imager (AOLI) which will be installed at the William Herschel Telescope and start operations in late 2013/early 2014. Additionally, J. Kotilainen presented a sample of QSOs ($z<0.5$) located in the SDSS Stripe 82, whose deep photometry allows for morphological studies of the host galaxies. A similar technique, but employing X-ray observations from different satellites, was used by E. Bottanici to create the Swift-INTEGRAL X-ray (SIX) survey. With a gain of two in sensitivity, he reported no evolution in the X-ray luminosity function of AGN. In that respect, A. Ermash proposed a new method for estimating the AGN luminosity function from the optical emission lines, which gives comparable results to those derived from the X-rays.

\bigskip

{\bf Galaxy mergers on all scales}\\
Session 5 was dedicated to galaxy mergers. One of the main topics was the incidence of AGN in merger systems,  particularly in luminous and ultra-luminous IR galaxies, as a key element to constrain galaxy-evolution theories. Another topic was the search for supermassive binary BHs, and the challenges faced when interpreting light curves and jet-structures to uniquely identify them.

Ultra luminous infrared galaxies (ULIRGs, i.e. with $L_{\rm IR(8-1000\mu  m)} > 10^{12}\,L_{\odot}$) and most of the local LIRGs ($L_{\rm IR} = 10^{11}-10^{12}\,L_{\odot}$)  are major mergers or strongly interacting systems. Only in galaxies with infrared luminosities lower than few $10^{11}\,L_{\odot}$  environmental or internal secular evolution becomes important. In the review on `AGN and star formation activity in local luminous and ultraluminous infrared galaxies', A. Alonso-Herrero revealed that 
the detection rate of AGN in local LIRGs is 60-65\%, close to that in ULIGRS (70\%), with no dependency on the infrared luminosity of the system. She explained that 
although the AGN contribution to the $L_{\rm IR}$ of the galaxy increases with the infrared luminosity, the fraction of non-optically detected AGN does as well, indicating that in high-$L_{\rm IR}$ systems the AGN is much more embedded.
In contrast, C. Villforth found no enhanced merger rates in AGN hosts compared to normal galaxies using a sample of local ($0.5<z<0.8$) X-ray selected AGN ($L_X\sim10^{41}-10^{44}\,{\rm erg\,s^{-1}}$) compared with matched control samples.  
S. K\"onig showed the spectacular case of the LIRG NGC 1614 ($L_{\rm IR}\sim 10^{11.6}\,L_{\odot}$), a LINER galaxy with enhanced star formation in a ring-like structure of radius $\sim 250$\,pc. From observations with the Sub-Millimeter Array, she found indications for the ring being fed with cold molecular gas from the dust lane and the tidal debris produced by a minor-merger event. L. Moser presented a sample of low-luminosity QSOs (LLQSOs) consisting of all Type 1 AGN with $z<0.06$ in the Hamburg/ESO QSO survey, and showed that in redshift, gas mass, and luminosity these objects lie in between the NUclei of GAlaxies (NUGA) and the Palomar Green (PG) QSO samples. G. Busch also showed that most of the LLQSOs hosts are late-type galaxies with a very high bar fraction ($\sim80$\%) and that 10\% of them are interacting systems. 

In the course of major merging events, binary supermassive BHs are expected to form. M. Mezcua presented evidences for binary BHs in X-shaped radio galaxies and double-nucleus ${\rm galaxies}$.  She argued that the low detection rate of such systems can be explained by the disruption of the accretion disks during the merging process. C. Villforth presented four double-peak [O{\sc iii}] sources with identified double-nuclei in infrared, radio and/or X-rays. First results from  optical spectroscopy reveal disturbed kinematics indicative for ongoing merging. On the other hand, S. Britzen proposed new methods for searching for binary BHs using long-term monitoring archive VLBI data. She explained that even though 
in the latest phases of BH merging the two individual cores can not be resolved by radio interferometry, perturbations in the jet-structure and distinct kinematics of the jet-components can be explained by the interaction of binary BHs.

\bigskip

{\bf Star formation and accretion at high redshifts}\\
Some of the key questions in the high redshift session were how AGN activity is triggered,  what are the characteristics of AGN hosts at high-$z$, and how does the relation between the AGN and the star formation in the host galaxy evolve with redshift. The reliability of black hole mass estimates and new method to explore the inner region of AGN using gravitational lensing were also discussed. 

In the review talk `Star formation and black hole accretion at high redshift', H. Netzer showed how
attempts to construct a `mass-sequence' of AGN (in analogy to that of star-forming galaxies)
are hampered by several biases of the different samples. He summarized it as: `We are living in the Era of the selection effects'. He also revealed that all sources detected by Herschel/SPIRE at $z\sim4.8$ are AGN dominated, and that 25\% of those have high star formation rates (${\rm SFR\sim2000-5000}\; M_{\odot}\,{\rm yr^{-1}}$). These are AGN with high luminosity and high-$M_{\rm BH}$. 
In contrast, low-to-medium luminosity AGN-hosts have SFRs similar to those of inactive galaxies, according to the Herschel/PACS results presented by D. Rosario. He also reported higher detection rates of AGN in star-forming hosts than in passive galaxies, indicating that the high incidence of AGN in the `Green Valley' might not imply quenching of the star formation by the AGN, but a natural consequence of the mass selection effects. 
Conversely, in a study of IR, optical, and X-ray selected samples from the COSMOS survey, M. Symeonidis found that Type 2 AGN (with $L_X >10^{42}\,{\rm erg\,s^{-1}}$) at $z<1.5$ are about three times more likely to reside in dusty star-forming galaxies than it would be expected serendipitously if black hole accretion and star formation were unrelated events.
A. Hern\'an-Caballero  reported on the study of stellar populations of intermediate redshift ($0.6<z<1.1$) X-ray selected AGN from the SHARDS survey. He found that, unlike in the local universe, U-V color and average age of the stellar populations of the AGN-hosts are comparable to those of normal galaxies of similar masses.
S.X. Wang showed that $\sim 17$\% of the sub-millimeter galaxies (SMGs) might host an AGN with $L_X \gtrsim 10^{43}\,{\rm erg\,s^{-1}}$, based on ALMA observations of X-ray luminous SMGs. In a similar sample (X-ray and sub-mm luminous QSOs at $z\sim2$) F. Carrera found covering factors of $0.4-0.9$ and star formation rates of $\sim 1000 M_{\odot}\; {\rm yr ^{-1}}$ from fitting their SEDs.

Clear signs of AGN feedback have been observed in nearby and high redshift quasars, but their importance in shaping the black hole mass -- host galaxy relations is still under debate. 
C. Collet reported on IFU observations of powerful radio galaxies at $z\sim2$. First results on the gas kinematics show high velocity dispersion in regions where the jet might interact with the gas, but no indication of high velocity outflows induced by the jet-ISM interaction.
Nevertheless, at $z\gtrsim 0.5$ the presence of Low ionization Broad Absorption Lines (LoBALs) in the quasar-spectra seems to be more common. T. Urrutia showed that, on dust-reddened LoBALs quasars at $0.9<z<1.8$, the strength  of these ionized outflows anti-correlates with the star formation in the host galaxies.
A.-L. Tsai compared normal and red QSOs at $0.6<z<2.0$, finding that the higher MIR emission observed in the latter ones might be associated with dustier/stronger star formation, or with different torus properties. 

Black hole mass estimations based on C{\sc iv} can be very convenient for objects at $z \gtrsim 2$. The usage of the C{\sc iv} line has been questioned in the past. However, K. Denny showed that adopting the second moment of the line profile as measurement of its width in high-quality data results in black hole masses that are consistent with those determined using H$\beta$ (in single epoch RM). She explained that the presence of a non-reverberating component arising mainly at the core of the C{\sc iv} line affects the FHWM measurements and can be the reason for previous discrepancies.
In agreement with this finding, D. Sluse showed that in the lensed quasar Q2237+0305, the size of the broad C{\sc iv} emitting region is about six times smaller than the region emitting narrow C{\sc iv}. These results were obtained using gravitational lensing techniques to systems where a high-$z$ AGN is behind a nearby galaxy. He explained that, in this configuration, the stars in the foreground galaxy act as lenses, magnifying areas in the quasar that are as large as approximately the BLR-size. 
A new method to improve the photometric redshift estimations for high-$z$ candidates using  spectral analysis of the X-ray data to complement the IR-to-UV information was proposed by J. Buchner.

\section{Open questions}
During the workshop, a number of questions were raised and discussed. Although most of them have being already discussed above, here we emphasize some of them together with additional comments from the participants of the Workshop. In particular, L. Foschini ${\rm [L.F.]}$, G. La Mura ${\rm [G.LM.]}$, and A. Ermash ${\rm [A.E.]}$ provided written answers, from which we extracted some excerpts.

\begin{enumerate}
\item {\it What have we learned from observations of the Galactic Center and nearby galaxies about accretion onto BHs, circumnuclear star formation, and their relation?} \newline

\hspace{-1.03cm} 
\begin{minipage}[b]{\textwidth}
The Galactic Center provides valuable information about star formation in the
vicinity of a supermassive black hole and possibly about processes of feeding and feedback.\\
${\rm [L.F.:]}$``SgrA* is perhaps the template of a low-luminosity AGN (Eckart's talk), but is
also giving now important insights on the jet formation or Galactic winds
(Fermi bubbles, Su et al. 2010) and
 the accretion of
matter onto a SMBH, as the incoming molecular cloud G2 will likely show in
the forthcoming years (Schartmann's talk)''.  \\
${\rm [G.LM.:]}$``It is
fundamental for understanding the influence of compact massive objects
on stellar and gas dynamics, including the possibility of gas cloud
disruption and possible black hole feeding.''\\
${\rm [A.E.:]}$``The question about formation of
Nuclear Star Clusters (NSC) is also intriguing (Capuzzo's talk) [...]
There are galaxies with NSC and with no BHs present (but they might be just below our 
current detection limit), and there are some galaxies where BH and NSC
are present simultaneously. [...] So, do very massive black holes somehow prevent formation of an NSC or are they
disrupted during the process of black hole growth?''\\

\end{minipage}

\hspace{-1.03cm} 
\begin{minipage}[b]{\textwidth}

However, to learn about AGN, the study of active nearby galaxies is fundamental.\\
${\rm [G.LM.:]}$``[...] the
power of nuclear activity may drive different effects on the
surrounding environment, starting from the structure of the accreting
flows (radiative efficiency, formation of a disk and of a broad line
region), up to the effect on the host galaxy''.
\end{minipage}

\item {\it Are Narrow Line Seyfert 1s (NLSy1s) a `special class' of AGN?} \newline

\hspace{-1.03cm} 
\begin{minipage}[b]{\textwidth}
In this respect there is no consensus among the participants. M. Valencia-S. argued that
NLSy1s constitute simply the lower $M_{\rm BH}$ end of the Type 1 population, while O. de Xivry showed
that secular processes are much more important in NLSy1 hosts than in Seyfert 1 galaxies.
The latter might be related with the different physical conditions in the narrow
line region of NLSy1s (Xu's talk). Other comments were:\\
${\rm [L.F.:]}$``NLSy1s are not AGN intrinsically different
from Seyfert 1s''. ``On the other side, yes,
NLSy1s are special, and specifically radio-loud NLSy1s [...] The common knowledge indicated that
jets in AGN required a threshold mass of the central black hole ($\sim10^8\,M_{\odot}$),
while jets in other sources do not have such a requirement [...]
The now confirmed presence of powerful relativistic jets in NLSy1s, which
in turn have low mass BHs removes this threshold, making thus possible to attempt a reliable
unification of jets at all scales [...] Therefore, just by renormalizing for the mass of the
central compact object, the populations of jets in Galactic binaries and AGN now merge
together.''\\
${\rm [A.E.:]}$``This issue should be addressed considering NLSy1s as a whole [...]
All host galaxies of NLSy1s have pseudobulges [...] But, judging by Sersic index,
host galaxies of broad line Seyfert 1s (BLSy1s) have both types of bulges [...]
The narrow line region, as was stressed by D. Xu in her talk, tend to be less dense in NLSy1s than in
BLSy1s [...] Also it was noted that the accretion disk in NLSy1s has a much higher temperature. And
the black holes themselves really seem to be undermassive ones, accreting at
high Eddington ratios. AGN, most likely, do not possess spherical symmetry so, inevitably, orientation
influences the properties of radiation that comes to observer.''\\
${\rm [G.LM.:]}$``I'd rather say yes. Of course it depends on the way we try to
define this class [...] Furthermore, it is even difficult to define a
set of general properties that NLSy1s universally share. However, [...] it can be stated
that NLSy1 galaxies result from nuclear activity powered by
systematically small mass BHs, accreting at rather high Eddington
ratios, with hints of high ionization plasma and
that all the mentioned details can by no means fit in a simple
geometrical interpretation. This makes NLSy1 galaxies undoubtedly peculiar''.
\end{minipage}

\item {\it How are jets and outflows related to the accretion radiative efficiency and the black hole spin?} \newline

\hspace{-1.03cm} 
\begin{minipage}[b]{\textwidth}
The BH-spin is expected to play an important role in the formation/launching of jets and/or as 
witness of the merging history (Ward's talk).  While current spin measurements are
still model dependent, new methods for estimating it were discussed (Done's talk).
The relation between the accretion rate and the feedback mechanisms (jet or outflows)
is not clear. On one hand, high accretion rates seem to be required for producing outflows
(Urrutia's talk), but on the other, several examples of low luminosity (and low accretion rate)
AGN with jet-dominated radio emission were also presented (e.g. Fernandez Ontiveros' talk).
\end{minipage}

\hspace{-1.03cm} 
\begin{minipage}[b]{\textwidth}

${\rm [L.F.:]}$``The radiative efficiency is still to be understood. 
The composite disc seems to be a valid structure, but there is need of
data to confirm this hypothesis. The BH spin is important, but is
not a fundamental parameter [...] The BH spin alone is not
a driver of the jet, but what is important is the slip between the angular
velocity of the central compact object and that of the magnetic field
threading the horizon (likely linked to the accretion disc). The value of
the slip factor could also explain the difference between radio-loud and
quiet AGN.''\\
${\rm [G.LM.:]}$``A combination
of physical parameters, like black hole mass, spin and accretion rate is
more likely to explain why jets and outflows can be set up or not.''\\
Another example of the apparent ambiguity.
${\rm [A.E.:]}$``from microquasar studies, these objects can be in two states:
soft-high and hard-low. 
In soft-high regime the accretion rate is high, radiative efficiency is also high
and no jet is present. 
In hard-low state the accretion rate is low, radiative efficiency is also low and
the jet is often present. 
In the NLSy1s, a jet is launched at high Eddington ratios, sometimes close to unity.''
\end{minipage}

\item {\it What is the structure of the broad line region (BLR) and how is it connected to the toroidal obscuration?} \newline

\hspace{-1.03cm} 
\begin{minipage}[b]{\textwidth}
${\rm [L.F.:]}$``The interesting talk by
Peterson indicated that there is
no unique signature of a single structure. Rather, the cases of 3C120 and
Mkn 335 suggest even different structures at different radii of the BLR.
The connection of the BLR with the toroidal obscuration seems to be
regulated by the AGN continuum as it interacts with a clumpy medium, and
with the molecular torus existing beyond a certain radius (of sublimation)
from the central mass black hole.''\\
${\rm [G.LM.:]}$``It seems positive that the BLR has a flattened
geometry, which is most often seen at small to moderate inclinations, [...] and that its motion
pattern is mostly under the dynamical influence of the central engine [...] 
 It is also widely accepted that the motion pattern is a
combination of rotation, turbulence and radial flows, possibly
involving different matter phases. Considerations on the conservation
of angular momentum and on the anisotropy of the radiation field make
it very likely that toroidal obscuration occurs preferentially on the
same plane where the BLR flattened structure lies.''
\end{minipage}

\item {\it How much can we trust the different black hole mass-estimation methods?} \newline

\hspace{-1.03cm} 
\begin{minipage}[b]{\textwidth}
One of the main sources of uncertainty in the black hole mass-estimations of Type 1 AGN
when using reverberation mapping (RM) is the factor $f$ that relates the `virial product'
with $M_{\rm BH}$. The revolutionary technique of velocity-resolved RM promises a better 
understanding of the geometric structure of the broad line region and thereby a better determination of $f$ (Peterson's talk).
\end{minipage}

\hspace{-1.03cm} 
\begin{minipage}[b]{\textwidth}
Using a mean value of $\langle f \rangle \sim 5$, how reliable are RM-based black hole masses?
``At the level of 0.3-0.4\,dex, they probably are reasonably good,
certainly for measurements based on the Balmer lines'' (B. Peterson, this proceedings).
Single epoch RM estimates of the black hole mass are statistically reliable at about the
same level, but this is not always the case in particular sources (Ilic's talk).\\
On the other hand, it seems that active and inactive galaxies follow the same
$M_{\rm BH} - \sigma_{\ast}$ correlation (Woo's talk) with no redshift evolution (Wisotzki's talk)
and therefore this can also be used for black hole mass-estimations.
\end{minipage}

\item {\it Do non-hidden broad line region AGN exist or is the lack of broad lines in these objects a matter of observational biases (e.g., spatial resolution, sensitivity)?} \newline

\hspace{-1.03cm} 
\begin{minipage}[b]{\textwidth}
The existence of AGN without a broad line region has been hypothesized in the last decade.
P.C. Yu presented them as a possible phase in the evolution of an AGN. Here some opinions
of other participants.\\
${\rm [L.F.:]}$``Rather than instrumental biases, I prefer to think to some
transient episodes of (partial) obscuration, when the source as observed
in different times can change its observational classification,
which is consistent with a clumpy nature of the molecular torus''.\\
${\rm [A.E.:]}$``There are objects in which we
do not see broad lines due to obscuration/orientation effects.  
But objects with lower accretion rates might switch to different accretion
mechanisms and do not possess a BLR. Such AGN have been observed.''\\
${\rm [G.LM.:]}$``My opinion is that objects actually lacking
a BLR exist [...] I do not think that the BLR can be identified with the
accretion disk. So, if the central accretion flow does not radiate
enough power, the conditions for a BLR dynamical stability may not set
up and the BLR would not form (or would quickly disappear) in such
objects.''
\end{minipage}

\item {\it What is the character of the AGN/star formation relation at low/high redshifts: coincidental, causal, self-regulated?} \newline

\hspace{-1.03cm} 
\begin{minipage}[b]{\textwidth}
${\rm [G.LM.:]}$``Many details depend on how galaxies evolve, because
the path followed by field galaxies and cluster members can be
appreciably different, mainly due to the frequency of interactions,
disturbances and mergers. [...] Our results and the other works
led me to the opinion that the most appropriate answer is a
self-regulated mechanism. When gas is perturbed, the first effect we
may expect to see is star formation. If the involved mass is small,
the process may end up with the first stars blowing away the
surrounding medium, and the tale is over with a dwarf stellar
system. If the involved mass is larger, gas can eventually be trapped
and feed a super massive central object. [...]
The time scale to bring appreciable amounts of fuel to
the appropriate distance is quite confidently longer than the required
time lag to trigger star formation there around. However, when nuclear
activity is taking place, a complex balance between enhancing and
quenching effects must again be accounted for. Very large mass systems
could be very quick in storing much fuel in the nucleus and trigger a
powerful AGN, which could sweep the host galaxy medium and set an
upper limit in the range of mass assembly. Intermediate systems could
be able to go through some interplay stages, in which AGN and star
forming activity trade mass and energy, until the circum-nuclear
regions of the galaxies run out of gas, because it has been accreted,
blown away or turned into long-living stars, so that the galaxy
becomes quiescent, at least until something happens that pushes again
appreciable amounts of gas from the outer regions towards the nucleus.''\\
${\rm [A.E.:]}$``
Starburst and AGN activity can, in principle, be coincidental if we do not speak
about star formation in the centers of galaxies. 
What happens in the whole galaxy may not be connected with events in its very
center.
As was stressed in the talk by C. Mundell, the main question about AGN
phenomenon is how to bring matter to the vicinity of the black hole.
It seems that complete quenching does not happen, maybe except in the most powerful
quasars at high redshifts.
Maybe the most correct answer is that AGN - {\it nuclear} star formation relation is
casual and self-regulating.''
\end{minipage}

\end{enumerate}

\section{Disclaimer}
The aim of this summary is to offer a compact resume of the main results presented at the workshop. The length of this text compared to the variety of subjects discussed during the Workshop does not allow an extensive (and proper) treatment of each topic. In any case, although the statements here are formulated to describe, with the best of our intentions, the studies presented at talks and posters in the meeting, they constitute a subjective interpretation of each contribution. The readers are encouraged to refer to the individual paper proceedings and references therein for more information about each topic.

\acknowledgments

The workshop has kindly been partly supported by the Deutsche Forschungsgemeinschaft via grant SFB~956, the Max-Planck-Institut f\"ur Radioastronomie, the Universit\"at zu K\"oln, and RadioNet.
We thank L. Foschini, G. La Mura, and A. Ermash for their contributions and comments to this Summary. We also would like to thank all the participants of the workshop. 
We thank very specially Ms. Beate Naunheim and the local organizing committee for their support regarding the logistics of the event. 
M.V-S. aknowledges the funding from the European Union Seventh Framework Programme (FP7/2007-2013) under grant agreement N$^o$312789. 
M.V. is member of the International Max-Planck Research School (IMPRS) for Astronomy and Astrophysics at the Universities of Cologne and Bonn and the MPIfR. 
We had fruitful discussions with members of the European Union funded COST Action MP0905: Black
Holes in a violent Universe and the COST Action MP1104: Polarization as a tool to study the Solar System and beyond.

\end{document}